\documentclass{ws-ijmpa}
\usepackage[super,compress]{cite}
\usepackage{graphicx}
\begin{document}
\markboth{Authors' Names}{Instructions for typing manuscripts (paper's title)}

%
\catchline{}{}{}{}{}
%

\title{Triple Gauge Couplings at Future Hadron and Lepton Colliders}

\author{Ligong Bian, Jing Shu, Yongchao Zhang}
\address{State Key Laboratory of Theoretical Physics and \\ Kavli Institute for Theoretical Physics China (KITPC), \\ Institute of Theoretical Physics, Chinese Academy of Sciences, Beijing 100190, P. R. China}

\maketitle


\begin{abstract}
  The $WW$ production is the primary channel to directly probe the triple gauge couplings. We analyze the $e^+ e^- \rightarrow W^+ W^-$ process at the proposed Circular Electron-Positron Collider (CEPC), and find that the anomalous triple gauge couplings and relevant dimension six operators can be probed up to the order of $10^{-4}$.  We also estimate constraints at the 14 TeV LHC, with both 300 fb$^{-1}$ and 3000 fb$^{-1}$ integrated luminosity from the leading lepton $p_T$ and azimuthal angle difference $\Delta \phi_{ll}$ in the di-lepton channel. The constrain is somewhat weaker, up to the order of $10^{-3}$. The limits on the triple gauge couplings are complementary to those on the electroweak precision observables and Higgs couplings.

\end{abstract}



\section{Introduction: Anomalous triple gauge couplings beyond the SM}

Given the 125 GeV Higgs has been found at the Large Hadron Collider (LHC), the standard model (SM) is in some sense complete. However, it might suffer from new physics, e.g. in the gauge sector, which can be examined with an unprecedented precision at future lepton colliders such as the Circular Electron-Positron Collider (CEPC) and LHC Run II. In this short note we summarize some of the key points in one recent paper by us~\cite{Bian:2015zha}, regarding the future prospects of anomalous charged triple gauge couplings (TGCs) beyond SM at future lepton and hadron colliders.

Due to non-Abelian nature of the weak interaction, there exist triple and quartic couplings among the EW gauge bosons in the SM. Here we focus only on the charged TGCs, i.e. those involving the couplings $WW\gamma$ and $WWZ$. With the anomalous contributions beyond SM, the charged TGCs can be generally parameterized as~\cite{Hagiwara},
\begin{eqnarray}
\label{eqn:lagrangianTGC}
&& {\cal L}_{\rm TGC}/g_{WWV} =
ig_{1,V} \Big( W^+_{\mu\nu}W^-_{\mu}V_{\nu} -W^-_{\mu\nu}W^+_{\mu}V_{\nu} \Big)
+ i\kappa_V W^+_\mu W^-_\nu V_{\mu\nu} \nonumber \\
&& \qquad + \frac{i\lambda_V}{M_W^2} W^+_{\lambda\mu} W^-_{\mu\nu} V_{\nu\lambda}
   + g_5^V \varepsilon_{\mu\nu\rho\sigma} \Big( W^+_{\mu} \overleftrightarrow{\partial}_\rho W_\nu \Big) V_\sigma \nonumber \\
&& \qquad - g_4^V W^+_\mu W^-_\nu \Big( \partial_\mu V_\nu + \partial_\nu V_\mu \Big)
 + i \tilde{\kappa}_V W^+_\mu W^-_\nu \tilde{V}_{\mu\nu}
+ \frac{i \tilde{\lambda}_V}{M_W^2} W^+_{\lambda\mu} W^-_{\mu\nu} \tilde{V}_{\nu\lambda} \,.
\end{eqnarray}
where $V = \gamma\,, Z$, the gauge couplings $g_{WW\gamma} = -e$, $g_{WWZ} = -e\cot\theta_W$ with $\cos\theta_W$ the weak mixing angle, the field strength tensor $F_{\mu\nu} \equiv \partial_\mu A_\nu - \partial_\nu A_\mu$ with $A = W\,, \gamma\,, Z$, and the conjugate tensor $\tilde{V}_{\mu\nu} = \frac12 \varepsilon_{\mu\nu\rho\sigma} V_{\rho\sigma}$, and $A \overleftrightarrow{\partial_\mu} B \equiv A ({\partial_\mu} B) - ({\partial_\mu} A) B$. Besides the SM TGCs, the Lagrangian Eq.~(\ref{eqn:lagrangianTGC}) contains 14 anomalous TGCs up to dimension six. The parity ($P$) and charge conjugate ($C$) conservative couplings beyond SM $\Delta g_{1,Z} \equiv g_{1,Z} - 1$, $\Delta \kappa_{\gamma,\,Z} \equiv \kappa_{\gamma,\,Z} - 1$ and $\lambda_{\gamma,\, Z}$ can be related to the effective field theory (EFT) beyond SM, e.g. the dimension-6 operators in the SILH basis~\cite{SILH,SILH2}
\begin{eqnarray}
\label{Lagrangian}
\Delta {\cal L} 
&=& \frac{i c_W\, g}{2M_W^2}\left( H^\dagger  \sigma^i \overleftrightarrow {D^\mu} H \right )( D^\nu  W_{\mu \nu})^i
+\frac{i c_{HW} \, g}{M_W^2}\, (D^\mu H)^\dagger \sigma^i (D^\nu H)W_{\mu \nu}^i \nonumber \\
&& +\frac{i c_{HB}\, g^\prime}{M_W^2}\, (D^\mu H)^\dagger (D^\nu H)B_{\mu \nu} \,
+\frac{c_{3W}\,  g}{6 M_W^2}\, \epsilon^{ijk} W_{\mu}^{i\, \nu} W_{\nu}^{j\, \rho} W_{\rho}^{k\, \mu} \,.
\end{eqnarray}
The $c_W$ operator is strictly constrained by precision measurements and can be neglected as a first order approximation,
with only three operators left at the Dim-6 level~\cite{operator1,operator2,operator3}
\begin{eqnarray}
\label{eqn:tgc2operator}
\Delta  g_{1,Z} &=&  - \frac{c_{HW} }{\cos^2\theta_W}   \,, \nonumber \\
\Delta \kappa_\gamma &=&  - ( c_{HW} + c_{HB} ) \,, \nonumber \\
\lambda_\gamma &=&  - c_{3W} \, ,
\end{eqnarray}
with $\Delta \kappa_Z = \Delta g_{1,Z} -  \tan^2\theta_W \Delta \kappa_\gamma$ and  $\lambda_\gamma = \lambda_Z$. Under such circumstance, the anomalous TGCs are related by the EW $SU(2)_L \times U(1)_Y$ gauge symmetry, and there is only three independent couplings in the $C$ and $P$ conserving sector.

\section{CEPC constraints}

At both lepton and hadron colliders the TGCs can be directly probed in  the $WW$ pair process. When the information of $W$ boson decay is taken into consideration, the kinematics of $e^+ e^- \rightarrow W^+ W^- \rightarrow f_1 \bar{f}_2 \bar{f}_3 f_4$ is dictated by five angles in the narrow $W$ width approximation: the scattering angle $\theta$ between $e^-$ and $W^-$ and the polar angles $\theta^\ast_{1,2}$ and the azimuthal angles $\phi^\ast_{1,2}$ for the decay products in the rest frame of $W^\mp$. The polarization of $W$ bosons can be described by the spin density matrix (SDM) which contains the full helicity information of the $W$ pairs. We resort alternatively to the differential cross sections with regard to the five kinematic angles, which are more physically intuitive, and examine the effects of anomalous TGCs on the distributions of final fermions. It is found that all the distributions of the five angles contribute significantly to the sensitives of TGCs in most of the channels.

To examine the response of the angular distributions to the TGCs, we expand the differential cross sections in terms of the aTGCs,
\begin{eqnarray}
\label{eqn:omegai}
\frac{{\rm d}\sigma}{{\rm d}\Omega_k} = \frac{{\rm d}\sigma_0}{{\rm d}\Omega_k} \Big[ 1 + \omega_i (\Omega_i) \alpha_i + \omega_{ij}  (\Omega_k) \alpha_i\alpha_j \Big] \,,
\end{eqnarray}
where $\Omega = \cos\theta$, $\cos\theta^\ast_{1,\,2}$, $\phi^\ast_{1,\,2}$ (or alternatively $\Omega = \cos\theta$, $\cos\theta^\ast_{\ell,\,\jmath}$, $\phi^\ast_{\ell,\,\jmath}$ with $\ell$ and $\jmath$ leptons and quark jets). It is straightforward to obtain analytically the linear coefficient functions $\omega_i$ from the differential cross sections, which are used by the LEP experimental groups to extract constraints on the anomalous couplings~\cite{Diehl:1993br}.

With a huge luminosity of 5 ab$^{-1}$ at CEPC at a center-of-mass energy of $\sqrt{s} = 240$ GeV, we can collect a total number of $8.6\times 10^7$ events of $W$ pairs, with 45\%, 44\% and 11\% decaying respectively in the hadronic, semileptonic and leptonic channels. With such a huge statistics, these anomalous TGCs are expected to be severely constrained. Here for simplicity we assume that the CEPC is optimistically designed such that the systematic errors are comparatively small and the TGC sensitivities are dominated by the statistical uncertainties. The SM radiative corrections are expected to be of the same order as the Dim-6 operators and cannot be simply ignored. Nevertheless, the high order corrections can be treated as a constant term and affect only the best fit values but not the relative errors. Since there is no measurement from CEPC yet, the radiative correction can be omitted for the time being.


For large numbers of events at CEPC, the statistical errors can be estimated to be $\sqrt{N_i}$ with $N_i$ the number of events. Then it is straightforward to estimate the sensitivities of the anomalous TGCs and the relevant Dim-6 operators in Eq.~(\ref{eqn:tgc2operator}). For the sake of simplicity we concentrate only on the semileptonic channel in this short note, where the ambiguities involving quark jets have been taken into account. More information from the pure leptonic and hadronic channels and the combined sensitivities can be found in Ref.~\refcite{Bian:2015zha}.

\begin{figure}[t]
  \centering
  \includegraphics[width=0.31\textwidth]{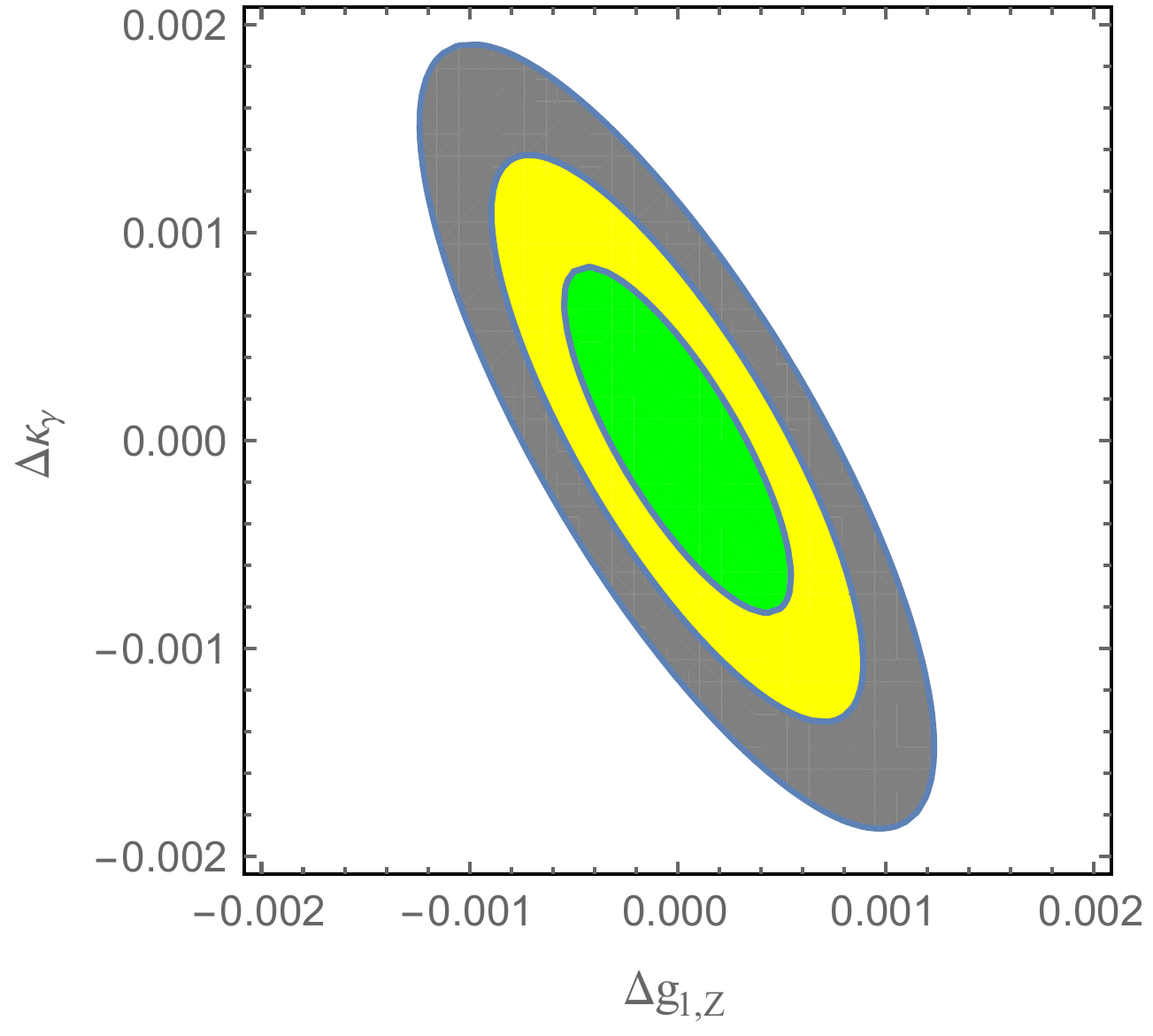}
  \includegraphics[width=0.3\textwidth]{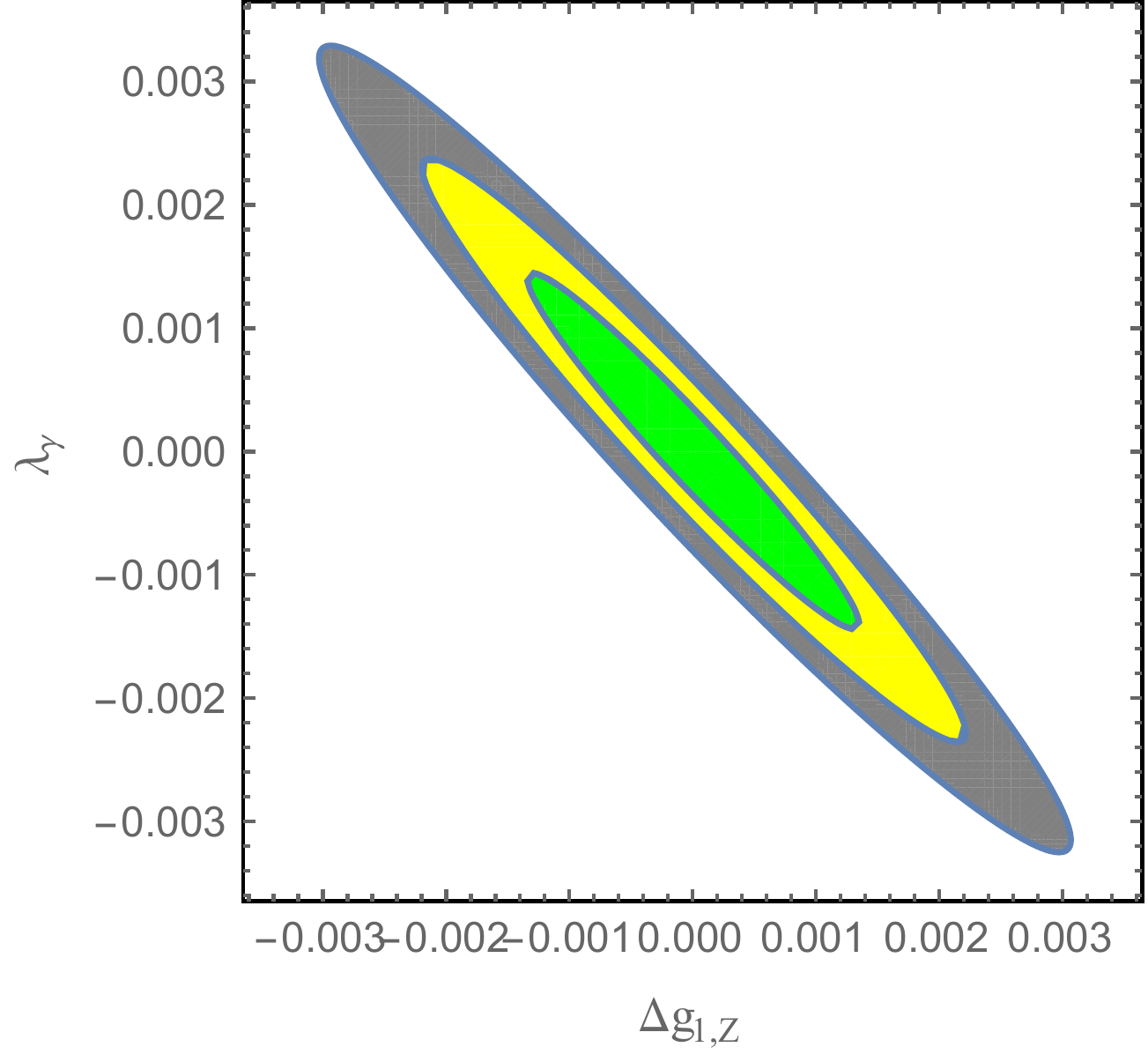}
  \includegraphics[width=0.31\textwidth]{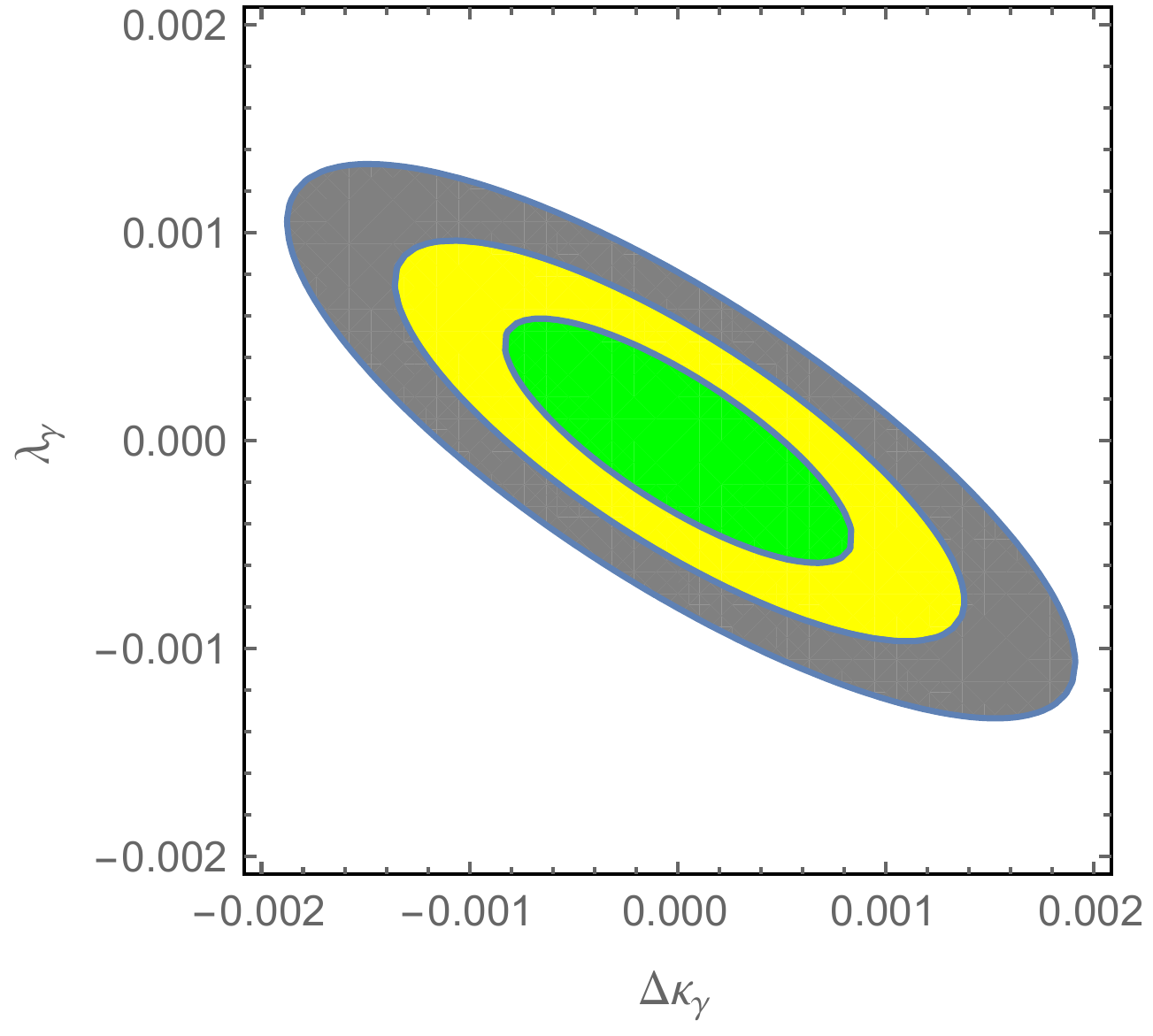} \\
  \includegraphics[width=0.31\textwidth]{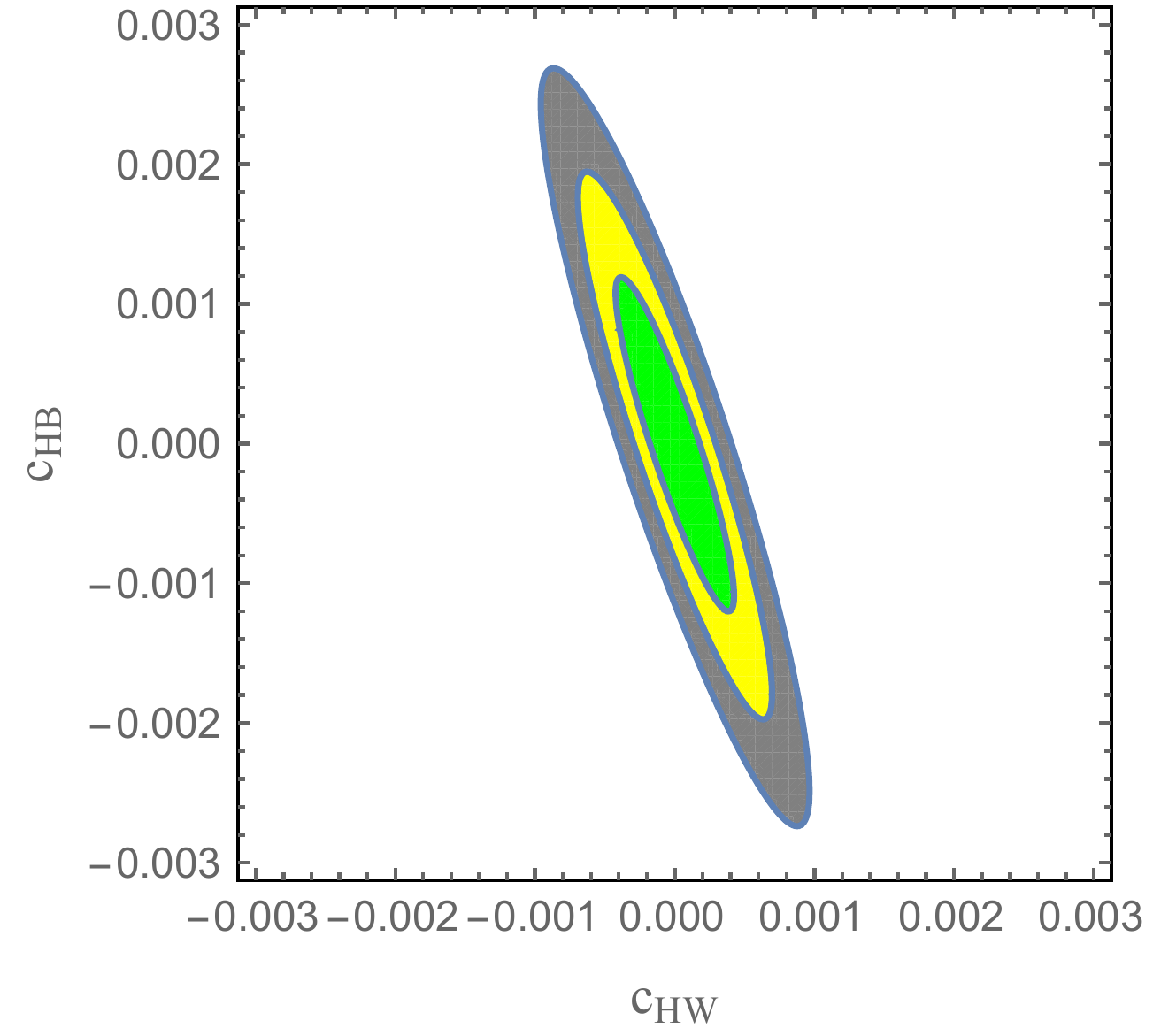}
  \includegraphics[width=0.31\textwidth]{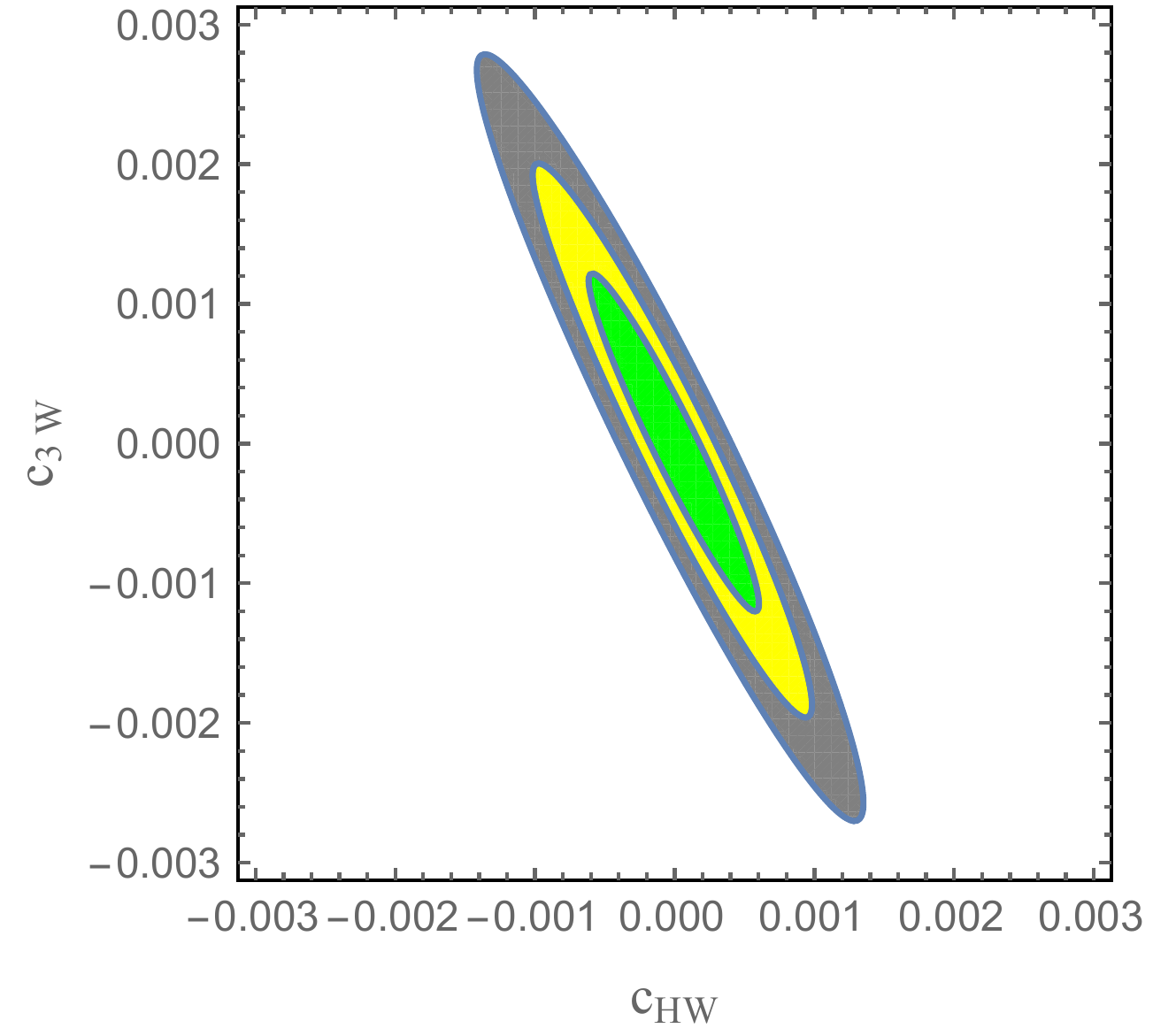}
  \includegraphics[width=0.31\textwidth]{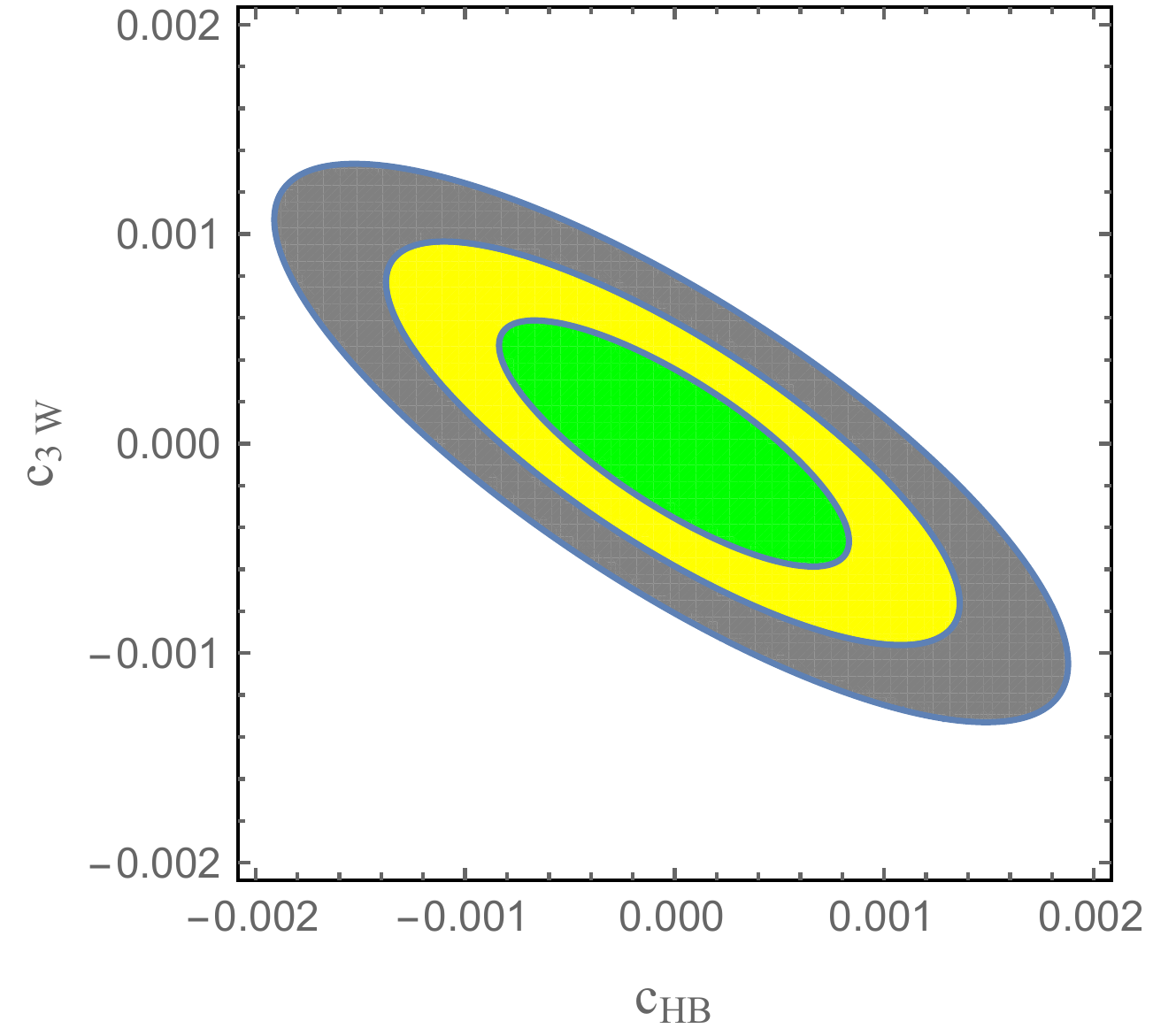}
  \caption{1$\sigma$ (green), $2\sigma$ (yellow) and $3\sigma$ (gray) allowed regions in the semileptonic channel for the aTGCs (upper panels) and Dim-6 operators (lower panels) at CEPC with $\sqrt{s} = 240$ GeV and an integrated luminosity of 5 ab$^{-1}$. In drawing the plots, two of the three couplings (or coefficients) are allowed to vary and the third one is fixed to zero.}
  \label{fig:contours}
\end{figure}

The expected semileptonic channel sensitivities of the aTGCs and Dim-6 operators at CEPC are presented in Table~\ref{tab:sensitivity}, where we assume an integrated luminosity of 5 ab$^{-1}$. All the sensitivities in this table are one parameter constraint where all other couplings or coefficients are fixed to zero. It is transparent that the limits on the TGCs and Dim-6 operators can reach up to the level of $10^{-4}$ in the semileptonic channel. When all the available channels are combined together, the sensitivities can be further improved. When two of the three anomalous couplings (or operator coefficients) are allowed to vary, we obtain the two-dimensional sensitivity plots presented in Fig.~\ref{fig:contours}. The correlations between the three TGCs and Dim-6 operators are, respectively,
\begin{eqnarray}
\label{eqn:rho}
\rho^{\rm TGC} = \left( \begin{matrix}
  1 && \\
  0.793 &  1 &  \\
  0.969 &  0.795 &  1
\end{matrix} \right) \,, \quad
\rho^{\rm operator} = \left( \begin{matrix}
  1 && \\
  0.906 &  1 & \\
  0.956 &  0.795 &  1
\end{matrix} \right) \,,
\end{eqnarray}
Obviously some of the TGCs (such as $\Delta g_{1,Z}$ and $\lambda_\gamma$) and operators are strongly correlated and there exist blind directions (such as $\Delta g_{1,Z} + \lambda_\gamma$) which are much less severely constrained. Such blind directions might be effected by experimental data and can be removed to some extent by incorporating the helicity information of $e^\pm$ and $W^\pm$~\cite{blinddirection,Falkowski}.

\begin{table}[t]
  \tbl{Estimations of the $1\sigma$ prospects (in units of $10^{-4}$) for the aTGCs and Dim-6 operators in the semileptonic decay channel of $WW$ process at CEPC with $\sqrt{s} = 240$ GeV and an integrated luminosity of 5 ab$^{-1}$.
  \label{tab:sensitivity}}
  {\begin{tabular}{@{}cccccccc@{}}
  \toprule
  $\Delta g_{1,Z}$ & $\Delta \kappa_\gamma$ &  $\lambda_\gamma$ &$\;$& $c_{HW}$ &  $c_{HB}$ & $c_{3W}$ \\
  \colrule
  2.19 & 3.33 & 2.35 &&  1.18 & 3.34 & 2.35 \\
  \botrule
  \end{tabular}}
\end{table}

\section{Constraints at LHC}

The TGCs can also be probed directly at hadron colliders in a similar process $q \bar{q} \to W^+ W^-$. These measurements are complementary to the EW precision tests, the accurate Higgs coupling probes, and all of these can be combined together to constrain the beyond SM physics~\cite{TGC:higgsdata,Ellis:2014jta}.
As a direct comparison, we consider simply the $WW$ production at the forthcoming LHC running at 14 TeV as an illustration. To suppress the huge QCD backgrounds, we focus on the purely leptonic decay channels $W \rightarrow e\nu, \, \mu\nu$. Though the neutrino events can not be fully reconstructed, the $p_T$ of charged leptons is widely used to study the TGCs. The azimuthal angle difference $\Delta \phi_{\ell\ell}$ projected onto the transverse plane in the lab frame, analogous to the azimuthal angles at lepton colliders, can be used to further improve the sensitivities. We do detector level simulations and apply the following cuts: for the charged leptons $l = e,\,\mu$, leading $p_T > 25$ GeV and subleading $p_T > 20$ GeV, $|\eta| < 2.5$, $\Delta R_{ll} > 0.4$, $m_{ll} > 15 (10)$ GeV, ${E}_T > 45 (15)$ GeV for the same (different) flavor channels, with the additional cut $\left| m_{ll} - M_Z \right| >15$ GeV for the same flavor channels. It is found that the anomalous couplings tend to generate large $p_T$ events. To optimise the constraints, we set further the leading $p_T > 300$ GeV and $> 500$ GeV respectively for a luminosity of 300 fb$^{-1}$ and 3000 fb$^{-1}$ and apply the cuts $\Delta \phi_{ll} > 170^\circ$,\footnote{When more data are collected, to improve the signal significance and suppress the SM background, the cuts can be made in a more aggressive manner.} with  the constrains collected in Table~\ref{tab:LHC}. For a luminosity of 300 fb$^{-1}$, the limits are of the order of magnitude of $10^{-3}$. When the luminosity is ten times larger, the constraints go two or three times stronger.

\begin{table}[t]
  \tbl{$1\sigma$ constraints on the TGCs and Dim-6 operators (in unit of $10^{-4}$) from the same flavor leptonic decay channels of $pp \rightarrow W^+W^-$ at 14 TeV LHC with a luminosity of 300 fb$^{-1}$ and 3000 fb$^{-1}$. \label{tab:LHC}}
  {\begin{tabular}{@{}cccccccc@{}}
  \toprule
  & $\Delta g_{1,\,Z}$   & $\Delta \kappa_\gamma$ & $\lambda_\gamma$ & $\;\;$ & $c_{HW}$ & $c_{HB}$ & $c_{3W}$  \\
  \colrule
  300 fb$^{-1}$   & 23   & 73  &  17    && 14   & 73   & 17  \\
  3000 fb$^{-1}$  & 11   & 30  & 5.7   && 6.3  & 30  & 5.7 \\
  \botrule
  \end{tabular}}
\end{table}

\begin{figure}[t]
  \centering
  \includegraphics[width=0.6\textwidth]{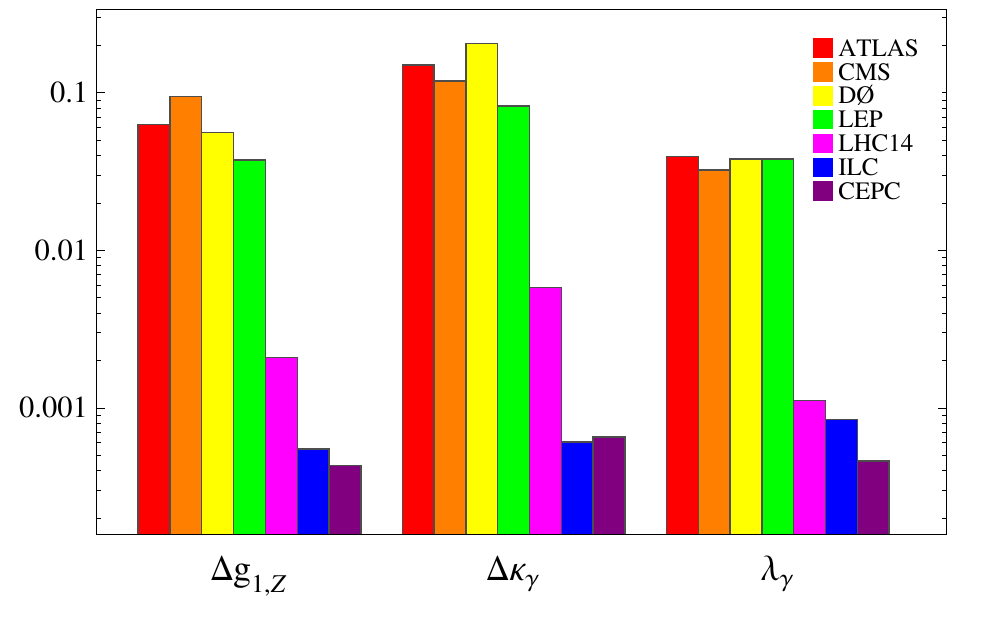}
  \vspace{-.5cm}
  \caption{Current and future 95\% confidence level constraints on the aTGCs. See text for details.}
  \label{fig:constraints}
\end{figure}

We collect in Fig.~\ref{fig:constraints} the current 95\% confidence level constraints on the anomalous TGCs $\Delta g_{1,\,Z}$, $\Delta \kappa_\gamma$, $\lambda_\gamma$  from LEP, Tevatron and LHC and that from 14 TeV LHC and the future lepton colliders CEPC and ILC. The current lepton and hadron collider bounds are from Ref.~\refcite{ATLAS5}, the data for LHC 14 TeV assume a luminosity of 3000 fb$^{-1}$, the limits for both CEPC and ILC use only the semileptonic channel, and for ILC they are the limits at $\sqrt{s} = 500$ GeV with a luminosity of 500 fb$^{-1}$ from Refs.~\refcite{ILC:TDR,AguilarSaavedra:2001rg}. It should be noted that at ILC the injected beams are polarized leptons, which is different from the circular collider CEPC. Comparing na\"ively the limits in this figure, the 14 TeV LHC and future lepton colliders can improve the limits on the aTGCs by one to two orders of magnitude. At future lepton colliders, using more decay channels, higher energies and larger luminosity can improve further the constraints in this figure.

\section{Conclusion}

In the precision era, the triple couplings among the SM electroweak gauge bosons is an essential part to test the SM in the gauge sector and set constraints on precision electroweak and Higgs physics, which can give us a powerful guidance on searching for new physics beyond the SM. 
$WW$ process is one of the most important channels at hadron and lepton colliders to measure directly the charged TGCs.  In this work we use the kinematical observable of decay products at CEPC and LHC to study the future constraints on the anomalous gauge couplings and the relevant three dimension-6 operators in the $C$ and $P$ conserving sector.  It is promising that constraints on the TGCs can be improved by two orders of magnitude and reach the order of magnitude of $10^{-4}$. A smaller gap between electroweak precision measurements, Higgs couplings and the TGCs will push us to reconsider the complementarity of them and the shrinking space for new physics.



\begin{thebibliography}{0}    

\bibitem{Bian:2015zha}
  L.~Bian, J.~Shu and Y.~Zhang,
  JHEP {\bf 1509}, 206 (2015)
  doi:10.1007/JHEP09(2015)206
  [arXiv:1507.02238 [hep-ph]].

\bibitem{Hagiwara}
  K.~Hagiwara, R.~D.~Peccei, D.~Zeppenfeld and K.~Hikasa,
  Nucl.\ Phys.\ B {\bf 282}, 253 (1987).

\bibitem{SILH}
  G.~F.~Giudice, C.~Grojean, A.~Pomarol and R.~Rattazzi,
  JHEP {\bf 0706}, 045 (2007)
  [hep-ph/0703164].
\bibitem{SILH2}
  R.~Contino, M.~Ghezzi, C.~Grojean, M.~Muhlleitner and M.~Spira,
  JHEP {\bf 1307}, 035 (2013)
  [arXiv:1303.3876 [hep-ph]].

\bibitem{operator1}
  A.~De Rujula, M.~B.~Gavela, P.~Hernandez and E.~Masso,
  Nucl.\ Phys.\ B {\bf 384}, 3 (1992).
\bibitem{operator2}
  K.~Hagiwara, S.~Ishihara, R.~Szalapski and D.~Zeppenfeld,
  Phys.\ Lett.\ B {\bf 283}, 353 (1992).
\bibitem{operator3}
  K.~Hagiwara, S.~Ishihara, R.~Szalapski and D.~Zeppenfeld,
  Phys.\ Rev.\ D {\bf 48}, 2182 (1993).

\bibitem{Diehl:1993br}
  M.~Diehl and O.~Nachtmann,
  Z.\ Phys.\ C {\bf 62}, 397 (1994).

\bibitem{blinddirection}
  G.~Brooijmans, R.~Contino, B.~Fuks, F.~Moortgat, P.~Richardson, S.~Sekmen, A.~Weiler and A.~Alloul {\it et al.},
  arXiv:1405.1617 [hep-ph].

\bibitem{Falkowski}
  A.~Falkowski and F.~Riva,
  JHEP {\bf 1502}, 039 (2015)
  [arXiv:1411.0669 [hep-ph]].

\bibitem{TGC:higgsdata}
  T.~Corbett, O.~J.~P.~\'{E}boli, J.~Gonzalez-Fraile and M.~C.~Gonzalez-Garcia,
  Phys.\ Rev.\ Lett.\  {\bf 111}, 011801 (2013)
  [arXiv:1304.1151 [hep-ph]].

\bibitem{Ellis:2014jta}
  J.~Ellis, V.~Sanz and T.~You,
  JHEP {\bf 1503}, 157 (2015)
  [arXiv:1410.7703 [hep-ph]].

\bibitem{ATLAS5}
  G.~Aad {\it et al.}  [ATLAS Collaboration],
  JHEP {\bf 1501}, 049 (2015)
  [arXiv:1410.7238 [hep-ex]].

\bibitem{ILC:TDR}
  H.~Baer, T.~Barklow, K.~Fujii, Y.~Gao, A.~Hoang, S.~Kanemura, J.~List and H.~E.~Logan {\it et al.},
  arXiv:1306.6352 [hep-ph].

\bibitem{AguilarSaavedra:2001rg}
  J.~A.~Aguilar-Saavedra {\it et al.} [ECFA/DESY LC Physics Working Group Collaboration],
  hep-ph/0106315.

\end{thebibliography}
\end{document}